\RequirePackage{fix-cm}
\documentclass[twocolumn]{svjour3}          
\usepackage{graphicx}
\usepackage{natbib}
\usepackage{amsmath, amsfonts, amssymb}
\usepackage{mathptmx}      
\usepackage[hyperindex,breaklinks]{hyperref} 
\usepackage{algpseudocode}

\usepackage{color}
\definecolor{vermilion}{rgb}{1,0.3,0}
\definecolor{venetianred}{rgb}{0.78,0.03,0.08}
\definecolor{tyrianpurple}{rgb}{0.4,0.01,0.24}

\newcommand{\eref}[1]{(\ref{#1})}
\newcommand{\aref}[1]{Appendix~\ref{#1}}
\newcommand{\fref}[1]{Figure~\ref{#1}}
\newcommand{\sref}[1]{Section~\ref{#1}}
\newcommand{\tref}[1]{Table~\ref{#1}}

\newcommand{\Kn}{K_{\mathrm{n}}}

\begin{document}

\title{Uniform random generation of large acyclic digraphs}



\author{Jack Kuipers \and Giusi Moffa}


\institute{Jack Kuipers \at 
Institut f\"ur theoretische Physik, Universit\"at Regensburg, \\
D-93040 Regensburg, Germany \\
\email{jack.kuipers@ur.de}
\and
Giusi Moffa \at
Institut f\"ur funktionelle Genomik, Universit\"at Regensburg, \\ 
Josef Engertstra\ss e 9, 93053 Regensburg, Germany \\
\email{giusi.moffa@ukr.de}
}

\date{}

\maketitle

\begin{abstract}
Directed acyclic graphs are the basic representation of the structure underlying Bayesian networks, which represent multivariate probability distributions. In many practical applications, such as the reverse engineering of gene regulatory networks, not only the estimation of model parameters but the reconstruction of the structure itself is of great interest. As well as for the assessment of different structure learning algorithms in simulation studies, a uniform sample from the space of directed acyclic graphs is required to evaluate the prevalence of certain structural features. Here we analyse how to sample acyclic digraphs uniformly at random through recursive enumeration, an approach previously thought too computationally involved. Based on complexity considerations, we discuss in particular how the enumeration directly provides an exact method, which avoids the convergence issues of the alternative Markov chain methods and is actually computationally much faster. The limiting behaviour of the distribution of acyclic digraphs then allows us to sample arbitrarily large graphs. Building on the ideas of recursive enumeration based sampling we also introduce a novel hybrid Markov chain with much faster convergence than current alternatives while still being easy to adapt to various restrictions. Finally we discuss how to include such restrictions in the combinatorial enumeration and the new hybrid Markov chain method for efficient uniform sampling of the corresponding graphs.
\end{abstract}

\section{Introduction}
\label{intro}

Acyclic digraphs or directed acyclic graphs (DAGs) comprise a collection of vertices or nodes linked by directional arrows. DAGs can be used to represent the structure of a popular class of probabilistic graphical models \citep{bk:Lauritzen96} known as Bayesian networks \citep{bk:Neapolitan2004}. These in turn represent probabilistic relationships between a large number of random variables, in short multivariate probability distributions, with the appealing property that conditional independence properties are encoded and can be read directly from the graph. They are widely used in many fields of applied statistics with especially important applications in biostatistics, such as the learning of epistatic relationships \citep{art:JiangNBV2011}. Triggered by the seminal paper of \cite{art:FriedmanLNP2000} Bayesian networks have become a popular tool for reverse engineering gene regulatory networks from large-scale expression data \citep{art:EStreibGAM2012}. Inferring the structure as well as the parameters of a model is particularly important to shed some light on the mechanisms driving biological processes. The estimation of DAGs or their equivalence class is a hard problem and methods for their efficient reconstruction from data is a very active field of research:  a recent review is given by \cite{art:DalySA2011} while some new methodological developments for estimating high dimensional sparse DAGs by constraint-based methods are discussed by \cite{art:ColomboMKR2012} and \cite{art:KalischB2007}. 

For simulation studies aimed at assessing the performance of learning algorithms which reconstruct a graph from data, it is crucial to be able to generate uniform samples from the space of DAGs so that any structure related bias is removed. The only currently available method relies on the construction of a Markov chain whose properties ensure that the limiting distribution is uniform over all DAGs with a given number of vertices $n$. The strategy is based on a well known idea first suggested by \cite{art:MadiganY95} as a Markov chain Monte Carlo (MCMC) scheme in the context of Bayesian graphical models to sample from the posterior distribution of graphs conditional on the data.  A specific algorithm for uniform sampling of DAGs was first provided by \cite{mdb01}, with the advantage over the standard MCMC scheme of not requiring the evaluation of the sampled graphs' neighbourhood, at the expense of slower convergence. The method was later extended \citep{ic02,icr04,mp04} to limit the sampling to restricted sets of DAGs.  An R implementation was also recently provided \citep{scutari10}. 

Since Markov chain based algorithms pose non-negligible convergence and computational issues, in practice random upper or lower triangular adjacency matrices are often sampled to generate random ensembles for simulation studies \citep[as for example implemented in the pcalg R package of][]{art:KalischMCMB2012}. Sampling over the space of triangular matrices however does not provide uniformly distributed graphs on the space of DAGs and is for example likely to perform poorly to obtain starting points for hill-climbing algorithms or slowly converging Markov chains. In fact due to the non-uniformity the risk of remaining within a small neighbourhood of certain graphs and more inefficiently exploring the space is increased. Likewise uniform sampling allows the correct evaluation of structure learning algorithms.  Finally, a uniform sample is essential when evaluating the prevalence of certain features in a population, in order to count the relative frequency of structural features in DAGs, which provides insight into the DAG landscape. This was recently pursued for example in \cite{bt13}, to assign causal or associative structural priors in search and score algorithms. 

In this work we therefore discuss a sampling strategy based on the recursive enumeration of DAGs but where no explicit listing is required and which follows directly from a combinatorial result of \cite{robinson70,robinson77}. An algorithm for generating uniform DAGs easily follows from the enumeration formula, which was also noted in an earlier technical report version \citep{mdb00} of \cite{mdb01}, where the method was thought computationally impractical and its complexity overestimated with respect to the Markov chain approach. Explicit evaluation of  both their complexities allows us to show that the enumeration method is actually the least expensive of the two. After re-establishing the practical convenience of enumeration based sampling our main contribution is twofold. First we exploit the asymptotic behaviour of DAGs to derive an extremely fast implementation for approximate (but highly accurate) sampling, which has the same (minimal) complexity as sampling triangular matrices. Very large DAGs can so be de facto sampled uniformly.

Taking a different direction we also devise a new hybrid MCMC method which is orders of magnitude faster than current MCMC methods but which shares their advantage of being easily adaptable to include restrictions or variations.  For example we can use this method to restrict the number of parents each vertex has or to weight the edges with a certain probability as is often performed when sampling triangular matrices.  The new hybrid MCMC method therefore offers a reasonably efficient alternative while avoiding bias by sampling from the correct space.

\section{Acyclic digraphs}

A directed graph or digraph on $n$ vertices has up to one directed edge between each pair of vertices, from node $i$ to node $j$ with $1\leq i,j\leq n$.  Acyclic digraphs or DAGs are those which admit no cycles so that there are no paths along the directed edges from a node to itself.  The total number $a_n$ of labelled DAGs with $n$ nodes, which is sequence A003024 in \cite{sloanes}, and its asymptotic behaviour are respectively \citep{robinson70,robinson73,stanley73}
\begin{equation} \label{totalDAG}
a_n=\sum_{k=1}^{n}(-1)^{k+1}\binom{n}{k}2^{k(n-k)}a_{n-k}, \qquad a_n\sim \frac{n!2^{L}}{Mq^n},
\end{equation}
with $L=\binom{n}{2}=n(n-1)/2$ and where $M=0.574\ldots$ and $q=1.48\ldots$ are constants.  The number of unlabelled DAGs was found later \citep{robinson77} and is recorded as sequence A003087 in \cite{sloanes}.

The directed edges of a DAG can be represented as entries with value 1 of a vertex adjacency matrix with remaining entries 0. Being acyclic the matrix can be made lower triangular by relabelling the nodes, and this observation lies behind the proof of \cite{mckayetal04} that the number of $n\times n$ (0,1)-matrices with only positive real eigenvalues is also $a_n$. Hence DAGs can be easily generated by sampling each of the $L$ elements of a lower triangular matrix from a Bernoulli distribution with probability $p=1/2$ and then permuting the $n$ labels. The graphs so obtained however are non-uniform on the space of DAGs since several permutations may correspond to the same graph.  For example, the DAG with no arcs (the 0 matrix) would be counted $n!$ times, while those with the most arcs (when all $L$ lower triangular elements are 1) only once. Choosing  $p \neq 1/2$ may reduce the overall non-uniformity but not remove it entirely, because the overcounting is sensitive to the structure of the DAG.

\section{Markov chain method} \label{markovchainmethod}

The algorithm of \cite{mdb01} starts with a given DAG, with no arcs say, and proceeds by repeatedly sampling an ordered pair of vertices from the $n$ available.  If an arc already joins the pair in that direction it is deleted, otherwise it is added as long as the graph remains acyclic.  Addition and deletion of an arc are inverse operations so that the resulting transition matrix $T$ is symmetric. As adding an arc to any pair $(i,i)$ would immediately create a cycle the probability to stay with the same graph is nonzero and at least $1/n$. One could exclude all but one such pair from the sampling to reduce this probability and marginally improve the convergence of the chain. Finally a path exists between any pair of DAGs since deleting all the arcs in turn leads to the empty DAG and the steps are reversible.  These three properties ensure that the stationary distribution of the Markov chain is uniform over the space of graphs and that an (approximately) uniform DAG can be generated by running the chain long enough.

The Markov chain method is very elegant, as no particular knowledge of DAGs is required apart from checking whether the graphs remain acyclic, which can be done in a time of the order between $n$ and the number of edges (typically around $n^2$) and on average in a time of order $n\log(n)$ \citep{ar78,mdb01}.  In fact when an edge is added, one only needs to check if cycles are added downstream which speeds up the calculation, but does not reduce its complexity. 

\subsection{Non-uniformity of the irreducible chain}\label{nonuniformitychain}

The main issue for the applicability of the method is how quickly the Markov chain converges.  The longest path is between a DAG with the most arcs ($L$) and the same graph with all the edges reversed. Therefore after $2L$ steps we obtain an irreducible transition matrix $U=T^{2L}$, whose elements though are far from uniform.  To illustrate this non-uniformity, we consider the extreme elements of $U$.  Moving along the longest path, consider first the transition probability of going from any DAG with $L$ arcs to the empty DAG with none.  Removing all the arcs in $L$ steps has a probability of $L!/n^{2L}$.  Adding arcs to arrive at the DAG with all the arcs reversed has the same probability due to the symmetry.  The transition probability of the single path moving through the empty DAG is then $(L!)^2/n^{4L}$.  Other paths exist as the $L$ additions and deletions can be ordered in $\binom{2L-2}{L-1}$ ways, with the first and the last step fixed as an addition and deletion respectively. The probability of moving along the other paths corresponding to different orderings is however lower. Hence the minimum transition probability, which is the minimum element of $U$, is 
\begin{equation} \label{Umineqn}
< \frac{(L!)^2}{n^{4L}} \binom{2L-2}{L-1} = \frac{L^2(2L-2)!}{n^{4L}} < \frac{(2L)!}{2 n^{4L}} ,
\end{equation}
which is actually $< 1/a_n^2$, much less than the required uniform value of $1/a_n$.  For comparison the ratio between the probabilities of the most and least likely DAGs, which can be interpreted as a measure of non-uniformity, deriving from sampling triangular matrices is only $n!$

This non-uniformity was overlooked in \cite{mdb01} leading them to underestimate the complexity of the Markov chain method.
To obtain an approximately uniform distribution the Markov chain needs to be run long enough (say $2Lk$ steps) so that the elements of the resulting transition matrix $U^k$ are close on a scale well below $1/a_n$.  This means we require $\vert (U^{k})_{ij} - 1/a_n \vert \ll 1/a_n$, $\forall ij$ or more accurately  
\begin{equation} \label{convergecondit}
\left[a_n \cdot (U^{k})_{ij}\right]^{\pm 1} -1 \ll 1 ,  
\end{equation}
where we take the positive power if the matrix element is larger than the uniform background and the negative power if it is smaller.  To see why we need such a restriction, consider the case for example where $\exists ij$: $[a_n \cdot (U^{k})_{ij}]^{\pm 1} \approx 2$ then the corresponding DAG will be sampled about twice or half as often as it should be and the sample cannot yet be considered uniform. This uniformity scale becomes finer as $n$ increases and leads to an additional dependence of $k$ on $n$ which increases the complexity of the algorithm.

\subsection{Speed of convergence and complexity}\label{markovspeed}

The difference between the maximum and minimum of the elements of $U^k$ decays at least exponentially $\sim\exp(-\alpha k)$ with a rate $\alpha$.  Proving lower bounds for the rate $\alpha$ is extremely hard and a generally known one, given by twice the minimum entry of $U$, is not enough to ensure that the algorithm is useful in practice since \eref{Umineqn} means that $k$ would need to be much larger than $a_n$ for convergence.  Instead, using the spectral decomposition as pursued in \aref{markovconvergence}, one can express a lower bound in terms of the largest eigenvalue of $U$ below that at 1.  More importantly, each element of $U$ must approach uniformity at least at this rate.   For the maximum element of $U$ for example we have
\begin{equation}
U^k_{\mathrm{max}}-\frac{1}{a_n} \leq C_{\mathrm{max}} \exp(-\alpha k)
\end{equation}
in terms of some constant $C_{\mathrm{max}}$ which is probably order 1 since the elements of $U$ must satisfy $0<U_{ij}<1$, and certainly less than $a_n$ as detailed in \aref{markovconvergence}.  The condition in \eref{convergecondit} is guaranteed to be satisfied if
\begin{equation} \label{maxconvcondit}
a_nC_{\mathrm{max}}\exp(-\alpha k) \ll 1 .
\end{equation}
Taking the logarithm in \eref{maxconvcondit}, it follows $\alpha k$ of order $n^2$ is sufficient to compensate for the quadratic growth of $\log(a_n)$. The more detailed consideration in \aref{markovconvergence} suggests that in fact $\alpha k$ has to be at least this order for convergence.

What matters next for the speed of convergence is how the eigenvalues of $U$ and $\alpha$ depend on $n$.  For DAGs with 2 or 3 nodes, $T$ can be easily filled out and we find $\alpha=0.575$ and $\alpha=0.594$ respectively.  By removing all but one of the arcs $(i,i)$ from the edge sampling, we can improve this to $\alpha=0.811$ and $\alpha=0.776$ instead. The behaviour should be verified, but as long as $\alpha$ does not decrease asymptotically with $n$, a uniformly sampled DAG can be obtained by throwing away the first $O(n^4)$ steps of the Markov chain. Analogously $O(n^4)$ steps should also be discarded between further samples to ensure they are independent. These conditions should be observed when using implementations like that in \cite{scutari10}.  As each step may involve checking for acyclicity, the entire algorithm then becomes approximately $O(n^5\log(n))$ which limits the size of graphs that can be sampled.

A marginal improvement in the speed of convergence could be achieved by starting each Markov chain on DAGs from a randomly drawn permuted triangular matrix.

\section{Enumeration method} \label{enumeration}

\begin{figure}
  \includegraphics[width=\columnwidth]{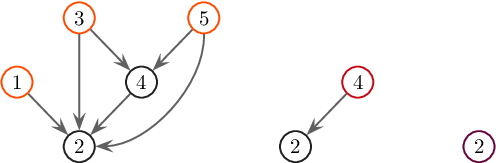}
\caption{Removing the outpoints from a DAG (and all their outgoing arcs), a smaller DAG is created until only outpoints remain.  For the DAG depicted on the left, one can first remove 3 outpoints, then 1 to be left with a single one.  Reversing the process by adding new outpoints and recording the number of possible ways of adding new edges and permuting the node labels leads to the recursive formula in \eref{ankrecur}.}
\label{fig:dagoutpoints}
\end{figure}

In order to obtain a uniform sample directly, we return to the enumeration of labelled DAGs detailed in \cite{robinson70,robinson77}.  Nodes with no incoming arcs are called `outpoints' and can be used to further classify and recursively enumerate DAGs. Due to the acyclicity each DAG has at least one outpoint. Let $a_{n,k}$ be the number of labelled DAGs with $n$ nodes of which $k$ are outpoints $(1\leq k\leq n)$.   Removing the latter and the arcs originating from them leaves a smaller DAG with $m=n-k$ nodes, with $s$ outpoints say. An example with $n=5$, $k=3$ and $s=1$ can be seen at the start of \fref{fig:dagoutpoints}. 

Reversing the process, the required DAGs can be built by adding $k$ new outpoints and allowing all the possible connections from them to the previous nodes.  For each of the $m-s$ non-outpoint nodes there are $2^k$ possibilities of having an arc or not.  Each of the $s$ old outpoints must be connected to at least once, giving a factor of ${2^k-1}$.  Finally the labels can be rearranged in $\binom{n}{k}$ ways, giving the recursions \citep{robinson70,robinson77}
\begin{equation} \label{ankrecur}
a_{n,k}=\binom{n}{k}b_{n,k},\qquad b_{n,k}=\sum_{s=1}^{m}(2^k-1)^{s}2^{k(m-s)}a_{m,s} , 
\end{equation}
with $a_{n,n}=1$ the DAGs with no arcs and $b_{n,k}$ auxiliary quantities useful later, also with the initialisation $b_{n,n}=1$.  The total number of DAGs with $n$ nodes is $a_n=\sum_{k=1}^{n}a_{n,k}$ which can be turned into the more compact form in \eref{totalDAG} by means of the inclusion-exclusion principle. The separation in terms of outpoints however is the key to sampling labelled DAGs uniformly.

\subsection{Recursively generating the outpoints}\label{outpoints}

The first step in the algorithm is to compute all the integers $a_{n,k}$ (and $b_{n,k}$) as well as the totals $a_n$.  From \eref{totalDAG} it follows that the number of bits needed to store them grows like $L$, while $2n+1$ integers are needed for each $n$ (since $1\leq k\leq n$).  Finally each of them is obtained by adding an order of $n$ smaller integers (multiplied by the binomial function) so that the process becomes $O(n^5)$ as detailed in \aref{complexity}.  However, they only need to be calculated once and then stored for sampling arbitrarily many DAGs.  Later we also discuss how to avoid their explicit evaluation.

The second step is, given $n$, to sample the number of outpoints. An integer is drawn between 1 and $a_n$, for example by sampling from a Bernoulli$(\frac{1}{2})$ the digits of a binary string long enough to represent $a_n$ (if the resulting number is larger than $a_n$ the string is rejected and resampled, which happens with probability less than a half). The chosen number of outpoints is the minimum $k$ for which the partial sum $\sum_{i=1}^ka_{n,i}$ is larger than or equal to the sampled integer.

The final DAG will be obtained by connecting the $k$ outpoints to a smaller one of size $m=n-k$, whose number of outpoints can be taken to be the minimum $s$ for which the partial sum $\sum_{i=1}^{s}(2^k-1)^{i}2^{k(m-i)}a_{m,i}$ is larger than or equal to a randomly drawn integer between 1 and $b_{n,k}$. Similarly a DAG with $m$ nodes and $s$ outpoints can be sampled by drawing an integer between 1 and $b_{m,s}$. Let $k_i$ denote the number of outpoints removed at each step, which can so be recursively sampled until their sum $\sum_{i=1}^{I} k_i =n$. In total one needs to sample and manipulate $I \leq n$ integers whose size in bits is at most of order $L$, so that generating the sequence is $O(n^3)$ as detailed in \aref{complexity}. 

Though for simplicity, we sampled an integer at each step above, the original one between 1 and $a_n$ could equally be reused to obtain the entire sequence of outpoints.  In fact, this can be computationally more efficient than resampling at each step and more directly follows the mapping between the integers and the DAGs.  For example, given the original integer $r$ sampled uniformly between 1 and $a_n$, we first find the minimal $k$ so that $1\leq r-\sum_{i=1}^{k-1}a_{n,i} \leq a_{n,k}$. The corresponding integer between 1 and $b_{n,k}$ is then
\begin{equation} \label{reusingr}
\left \lceil \frac{r-\sum_{i=1}^{k-1}a_{n,i}}{\binom{n}{k}} \right \rceil ,
\end{equation}
where we round up to the next integer.  As the numerator was already calculated when finding $k$ we can see why this can be cheaper than resampling.  Repeating these steps with the next partial sums gives the full sequence of outpoints.

\subsection{Reconstructing the DAG} \label{reconstruction}

To represent the sampled DAG we fill a null $n\times n$ adjacency matrix from the top left to obtain a lower triangular matrix. Since the recursion ends when all the remaining nodes are outpoints, the first $k_I\times k_I$ elements stay zero. These must connect to at least one of $k_{I-1}$ newly added outpoints which must not connect to each other, hence the elements of each column from 1 to $k_I$ in rows $k_I+1$ to $k_I+k_{I-1}$ can be sampled from a Bernoulli$(\frac{1}{2})$. Any samples where all the elements in a column are 0 can be rejected columnwise (alternatively a direct binomial sampling with this case excluded can be performed). 

When adding an additional $k_{I-2}$ nodes, the elements in columns 1 to $k_I$ and rows $k_I+k_{I-1}+1$ to $k_I+k_{I-1}+k_{I-2}$ are again sampled from a Bernoulli$(\frac{1}{2})$, while for each column $k_I+1$ to $k_I+k_{I-1}$ we need to ensure that not all new entries are 0.  Completing the adjacency matrix is a step of order $n^2$ since up to $L$ integers need to be sampled. The final sampled DAG can be obtained by permuting the labels only at the end, rather than at each step as suggested by \eref{ankrecur}, since each permutation must be equally likely.  Although different permutations may still correspond to the same DAG, the procedure of \sref{outpoints} for drawing the number of outpoints at each step is weighted so that the resulting sample is perfectly uniform.

The procedure is simply a 1 to 1 mapping between the integers 1 to $a_n$ and the DAGs with $n$ nodes.  In principle the originally sampled integer $r$ uniquely identifies the edges and node permutations.  For example the remainder when dividing by $\binom{n}{k}$ in \eref{reusingr} can be mapped to the permutation of the $k$ nodes.  It is simpler to ignore the remainders and their mappings while drawing the sequence of outpoints and just sample the edges and permutation at the end.  Since each set of compatible edges and node permutations is equally likely given the sequence of outpoints, the sample remains uniform.   The recursions weight the choice of the number of outpoints removed at each step so that the probability of selecting each sequence is proportional to the total number of corresponding DAGs.

A detailed sampling procedure is provided in \aref{pseudocode} in the form of pseudocode.

\subsection{Illustration of recursive sampling}\label{sampexample}

\begin{table}
\caption{\label{a5ktable}The number of DAGs with 5 nodes and $k$ outpoints.}
\begin{tabular}{l | c c c c c}
\hline\noalign{\smallskip}
$k$ & 1 & 2 & 3 & 4 & 5 \\
$a_{5,k}$ & 16\,885& 10\,710& 1\,610& 75& 1\\
\noalign{\smallskip}\hline
\end{tabular}
\end{table}

The method is best illustrate with an example.  For $n=5$, \tref{a5ktable} lists the quantities $a_{5,k}$.  An integer up to $a_5=29\,281$ is drawn uniformly, say $r=28\,405$.  Since $r$ is larger than $a_{5,1}$ this is subtracted to obtain $11\,520$, bigger in turn than $a_{5,2}$.  A further subtraction leaves $810$, smaller than or equal to $a_{5,3}$.  The desired DAG is the 810th with 3 outpoints out of 5.  Dividing by the $\binom{5}{3}=10$ ways of permuting the new outpoint labels means that the sampled DAG corresponds to the 81st of a list of $b_{5,3}=161$.  Also from \eref{ankrecur}
\begin{equation}
b_{5,3} = 56 a_{2,1} + 49 a_{2,2} ,
\end{equation}
where $a_{2,1}=2$ and $a_{2,2}=1$.  Since $81\leq 56\times 2$, a DAG with one outpoint is selected. In fact given that $a_{2,1}=2b_{2,1}$ and $\lceil 81/112\rceil=1$, the desired DAG is the first and only one of $b_{2,1}$.  Removing the outpoint, a single node remains and the sampled sequence is
\begin{equation} \label{examplesequence}
 k_1 = 3 , \quad k_2 = 1, \quad k_3 = 1.
\end{equation}
Sampling any of the integers between $27\,596$ and $28\,715$ would result in the same sequence of outpoints.

\begin{figure*}
  \includegraphics[width=\textwidth]{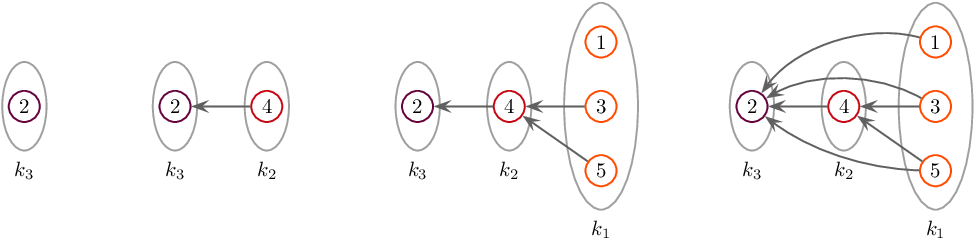}
\caption{Once a sequence of outpoints has been drawn, for example as in \eref{examplesequence}, the edges can be sampled by recursively adding new outpoints.  Here the labelling of the nodes respects the sampled permutation in \eref{examplepermutation}.  Starting with one outpoint from $k_3$, when the next one from $k_2$ is added an arc must also be included.  At least one arc must arrive at node 4 from the last three nodes.  Here there happened to be two.  So far the bold elements in \eref{exampletrimatrix} have been sampled.  Elements below the bold ones in the matrix in \eref{exampletrimatrix} may be drawn without restriction so that node 2 can receive an arbitrary number of arcs from nodes 1, 3 and 5; here all three were sampled.  The final sampled DAG is identical to the one at the start of \fref{fig:dagoutpoints}, merely organised by the sequence of outpoints.}
\label{fig:dagsampling}
\end{figure*}

To avoid confusion in \fref{fig:dagsampling}, the permutation of the node labels is drawn before reconstructing the edges.  Let the permutation be
\begin{equation} \label{examplepermutation}
\pi = \left(\begin{array}{ccccc}1 & 2 & 3 & 4 & 5 \\
             2 & 4 & 1 & 3 & 5
            \end{array}\right) ,
\end{equation}
in the two-line notation. Next a corresponding lower triangular adjacency matrix can be sampled
\begin{equation}\label{exampletrimatrix}
\begin{array}{c|ccccc}  & \;\;2\; & \;4\; & \;1\; & \;3\; & \;5\; \\
\hline
\;2\; & 0 & 0 & 0 & 0 & 0\\
\;4\; & \vec{1} & 0 & 0 & 0 & 0\\
\;1\; & 1 & \vec{0} & 0 & 0 & 0\\
\;3\; & 1 & \vec{1} & 0 & 0 & 0\\
\;5\; & 1 & \vec{1} & 0 & 0 & 0
            \end{array} .
\end{equation}
Let the $5\times5$ matrix inside the solid lines be denoted by $Q$.  The lower elements can be drawn uniformly as either 1 or 0, with certain restrictions.  First, not all the elements in bold in each column can be 0.  They correspond to possible connections between the old and new outpoints at adjacent steps in the sequence of outpoints.  Graphically the sampling of the matrix in \eref{exampletrimatrix} can be represented by the recursive addition of outpoints as in \fref{fig:dagsampling}.

The starting point is $k_3=1$ outpoint labelled by 2 because of the permutation in \eref{examplepermutation}.  Then $k_2=1$ new outpoint labelled by 4 is added. Since node 2 must stop being an outpoint it has to receive an arc from node 4 and the corresponding element $Q_{2,1}=1$.  At the next step the remaining $k_1=3$ are added ensuring that node 4 receives at least one arc.  Here it happens to receive 2 corresponding to the second column of $Q$ in \eref{exampletrimatrix}.

The second restriction is that the lower triangular matrix elements above, or to the right, of the bold ones must all be 0 since no arcs are allowed between outpoints added at the same time.  The remaining lower triangular matrix entries below the bold ones can be sampled without restriction.  In this case they correspond to arcs from nodes 1, 3 and 5 to node 2.  In the example they were all drawn to arrive at the final sampled DAG in \fref{fig:dagsampling}, which is just a redrawing of the first DAG in \fref{fig:dagoutpoints}.

To complete the example, the lower triangular matrix in \eref{exampletrimatrix} can be rearranged to standard ordering using $R_{\pi(m),\pi(l)}=Q_{m,l}$ leading to the adjacency matrix
\begin{equation}\label{exampleadjmatrix}
R = \left(\begin{array}{ccccc}  
0 & 1 & 0 & 0 & 0\\
\;0\; & \;0\; & \;0\; & \;0\; & \;0\;\\
0 & 1 & 0 & 1 & 0\\
0 & 1 & 0 & 0 & 0\\
0 & 1 & 0 & 1 & 0
\end{array}\right) ,
\end{equation}
which is also the adjacency matrix of the first DAG in \fref{fig:dagoutpoints}.

\subsection{Computational time and space complexity} \label{enumerationcomplexity}

As the different parts of the algorithm are performed sequentially, the complexity is determined by the first step, which is the most expensive. A uniform DAG can therefore be sampled from scratch in $O(n^5)$, marginally better than the $O(n^5 \log n)$ required by the Markov chain algorithm. Once the integers in \eref{ankrecur} have been computed and stored, each subsequent perfectly uniformly sampled DAG is obtained in only $O(n^3)$.  This is significantly better than the Markov chain algorithm which requires the same $O(n^5\log(n))$ for each subsequent approximately independent and uniform DAG. These considerations are important since in most practical cases the aim is to draw a reasonably large sample of DAGs, rather than a single one.

Not only the time complexity is improved, but the computational time is also vastly reduced.  In a simple Maple implementation, the first step is rather quick to perform, for example taking under a second for $n=100$.  When reusing the original integer drawn between 1 and $a_n$ it takes about a tenth of a second to sample each sequence of outpoints.  The final DAG given the sequence is then sampled in less than a twentieth of a second so that the total time for all the steps is still under a second.  At least 6 uniformly distributed DAGs can be obtained in a second before optimising the code further.  For reference, the computations were all run on a single core of a standard Intel Q9550 processor with 4GB of RAM.

The R implementation in \cite{scutari10} of the Markov chain method can perform about 80\,000 steps of the chain per second on the same machine.  This means that in the time it takes to sample each DAG perfectly uniformly using the recursive enumeration, the Markov chain only just becomes irreducible (recall that this takes $2L\approx10^{4}$ steps). Before it can be reasonably expected that an approximately uniform sample is provided by the chain it needs to run for hundreds of millions of steps ($100^4$) as discussed in \sref{markovchainmethod} and is therefore tens of thousands of times slower than the enumeration method for $n=100$.  This means going from the order of seconds to the order of hours and given the complexity arguments, things only get worse as $n$ increases.

The time complexity analysis of the first step of the enumeration algorithm, on the other hand, does not account for the the space required to store and access all the numbers $a_{n,k}$ and $b_{n,k}$ which together require up to about $n^4/4$ bits.  Even with the gigabits of fast memory available on modern computers, the space limit is reached when $n$ gets to several hundred.  In fact, as we will see in the next section, fast access is only needed for low $k$ reducing the order of bits to $n^3$. The memory space limit is then only reached by DAGs on about a thousand nodes but the time complexity still makes the sampling on such ultra-large spaces unfeasible.

The exact sampling is just a way of mapping between the integers between 1 and $a_n$ and DAGs, whose superexponential growth as in \eref{totalDAG} lies behind the space and time problems for large $n$.  As an indication of this growth, sampling a DAG with just 21 nodes is akin to sampling an atom uniformly from the observable universe.  To move to arbitrarily large DAGs, however, we can introduce an approximate method which is both extremely fast and highly accurate.  

\section{Arbitrarily large DAGs}\label{largesampling}

\begin{table}
\caption{\label{Aktable}The relative occurrence of the number of outpoints in large DAGs, $A_k$ multiplied by $10^{10}$.  $A_8$ is approximately $2.2\times10^{-12}$, so $k>7$ can be excluded at this level of accuracy.}
\begin{tabular}{lr@{ }r@{ }r@{ }r c c lr@{ }r@{ }r c c lr}
\hline\noalign{\smallskip}
$A_1$ & 5&743&623&733 & & & $A_4$ & 29&023&072 & & & $A_7$ & 15 \\
$A_2$ & 3&662&136&732 & & & $A_5$ &   &566&517 & & & $A_8$ &  0\\
$A_3$ &  &564&645&435 & & & $A_6$ &   &  4&496 & & &       & \\
\noalign{\smallskip}\hline
\end{tabular}
\end{table}

The number of outpoints sampled at each step is small and most often just 1.  In fact the fraction of DAGs with $k$ outpoints among $n$,
\begin{equation}
A_k=\lim_{n\to\infty} \frac{a_{n,k}}{a_n} ,
\end{equation}
converges as was proved in \cite{liskovets76} and to within $10^{-10}$ by $n=20$.  For $n>20$, within the uniformity limits of a 32 bit integer sampler, $k$ could be directly sampled from the limiting distribution in \tref{Aktable}.  Given the initial $k$ the next value $s$ in the sequence of outpoints can be drawn from the limiting conditional probabilities
\begin{equation} \label{limitcondprobs}
B_{s\mid k}=\left(1-\frac{1}{2^k}\right)^{s}\frac{A_{s}}{Z_k}, \qquad Z_k = \sum_s \left(1-\frac{1}{2^k}\right)^{s} A_{s} ,
\end{equation}
which follow from \eref{ankrecur} as detailed in \aref{limitcondprobsderivation}.  The $k$-dependent weight further shifts the distribution towards low $s$ so it would be computationally meaningful to store the small number of relevant probabilities.  The procedure is iterated until 20 nodes or fewer are left to be sampled,  when the tabulated distributions are no longer accurate enough. For $n\leq20$ the exact distributions can be tabulated or one may simply return to the exact enumeration method.

Generating the sequence of outpoints exploiting the limiting behaviour is of order $n$ while reconstructing the DAG remains order $n^2$.  Again these steps are sequential so that a uniform (at scales above $10^{-10}$) DAG can be drawn in $O(n^2)$ allowing the sampling of very large graphs. Sampling a lower or upper triangular matrix of the same size has the same complexity, which is therefore the minimum possible for sampling DAGs. As it is built on the underlying space of DAGs, the procedure above has the advantage over sampling triangular matrices of removing (most of) the bias, while being faster than other potentially unbiased alternatives such as the Markov chain method of \cite{mdb01}. 

For the latter, convergence of the transition matrix to a given level of accuracy is reached in $O(n^2)$ steps so each sample can be drawn in $O(n^3\log(n))$, as follows from the discussion in \sref{markovchainmethod}.  The probabilities with which different DAGs are sampled however vary wildly, with a more pronounced variability than observed when sampling triangular matrices. In the approximated method above instead, a zero weight is given to the rare event of drawing more than 7 outpoints at any step, resulting in a zero probability for all DAGs with any $k_i>7$.  The rest however are sampled exactly uniformly and again much faster than with the Markov chain method.

The accuracy of the convergence to the limiting distributions of $A_k$ nearly exactly doubles when doubling $n$.  Uniformity on the scale of a 64 bit integer sampler can be achieved by a more accurately calculated distribution of $A_k$ for $n>40$ while returning to the exact enumeration for $n\leq40$.

\section{Hybrid MCMC method} \label{hybridMCMC}

Although the approximate method of \sref{largesampling} will in practice give indistinguishable results from the full enumeration sampler, it has the philosophical problem that certain DAGs will never be sampled.  As an alternative, and one which allows restrictions and variations to be incorporated more easily, we propose a hybrid method combining the ideas from the recursive enumeration and Markov chain approaches.

In the enumeration based sampler, most of the effort goes towards sampling the sequence of outpoints $k_i$ added at each step.  As $\sum_{i=1}^{I}k_i =n$, this sequence is just an ordered partition of the number $n$.  A MCMC on the space of partitions can be introduced, where each partition is scored according to the number of DAGs corresponding to it.  Denoting each partition by a sequence in square brackets $[k_I,\ldots,k_1]$, to respect the order in \sref{enumeration}, and setting $a_{[k_I,\ldots,k_1]}$ to be the corresponding number of DAGs, from \eref{ankrecur} we have
\begin{equation} \label{partitiondagnumbers}
a_{[k_I,\ldots,k_1]} = \frac{n!}{k_I!\ldots k_1!} \prod_{i=1}^{I-1} (2^{k_i}-1)^{k_{i+1}}\prod_{i=2}^{I-1} 2^{S_{i-1}k_{i+1}},
\end{equation}
with $S_{i} = \sum_{j=1}^{i}k_i$. The first term accounts for the number of ways of organising the labels, the second for the number of ways of connecting edges between adjacent partition elements and the last term for the number of ways of connecting edges between non-adjacent partition elements.  Empty products are given the value of 1 so that $a_{[n]}=1$ counts the empty DAGs with no edges.

\subsection{Partitions and binary sequences}

There are $2^{n-1}$ ordered partitions of the integer $n$ which can be represented by the binary sequences of length $(n-1)$.  Each 0 corresponds to starting a new partition element whose size is one larger than the number of 1's since the previous 0.  For example the binary sequence $(0,1,1,0,0,1)$ gives the partition $[1,3,1,2]$ of $7$ when treating the brackets as zeros.  A Markov chain on the space of partitions can be built by flipping one of the elements of the binary sequence.  Changing a 1 to a 0 splits a partition element into two, while changing a 0 to a 1 joins two adjacent partition elements.  To introduce some probability of staying still, and ensure aperiodicity, the binary sequence can be extended with an artificial $n$-th element which does not affect the partition.  At each step of the chain one of the $n$ elements is uniformly drawn and flipped.  Each partition then has $(n-1)$ neighbours excluding itself and the chain takes $(n-1)$ steps to become irreducible.

The Metropolis-Hastings acceptance probability to move from partition $P$ to a proposed partition $\hat{P}$ is
\begin{equation}\label{accratio}
\min\left(1,\frac{a_{\hat{P}}}{a_P}\right) ,
\end{equation}
which involves the number of DAGs  in each partition from \eref{partitiondagnumbers}. The resulting MCMC scheme converges to a stationary distribution where the probability of each partition is proportional to the number of DAGs belonging to it.

Only one digit of the binary sequence is flipped at each step. A single partition element is then split or two neighbouring ones are joined depending on the direction of the flip. The partition elements directly involved in the modifications and their neighbours are the only ones which can be affected by changes in the edge configuration. Hence the terms corresponding to the unaffected partition elements simplify in the ratio $a_{\hat{P}}/a_P$. For example, given a partition 
\begin{equation}
P=[k_I,\ldots,k_{i+1},k_i,k_{i-1},\ldots,k_{1}] ,
\end{equation}
if the element $k_i$ is split into elements $c$ and $k_i-c$ to propose partition 
\begin{equation}
\hat{P}=[k_I,\ldots,k_{i+1},c,k_i-c,k_{i-1},\ldots,k_{1}] ,
\end{equation}
the ratio in the acceptance probability becomes
\begin{equation} \label{MCMCacceptratio}
\frac{a_{\hat{P}}}{a_P} = \binom{k_i}{c}\left(\frac{1-2^{-c}}{1-2^{-k_i}}\right)^{k_{i+1}}\left(\frac{1-2^{c-k_i}}{1-2^{-k_{i-1}} \mathrm{I}_{\{ i>1 \} }  }\right)^{c} 2^{c(k_i-c)} ,
\end{equation}
depending just on $k_{i-1},k_{i},c$ and $k_{i+1}$. For convenience $k_i$ is set to $0$ for $i$ outside $1, \ldots, I$ and an indicator function is included to correct for the case when $i=1$.

\subsection{Complexity of the MCMC steps}

To evaluate the complexity of calculating the acceptance ratio and sampling accordingly, the relative weight of different partitions is first taken into account.  In general, partitions with a larger number of smaller partition elements are more likely as they allow more DAGs. Therefore the chain will have a preference to move to and spend more time on them.  In fact the most likely partition is the one with all $k_i=1$ for which the calculation of the ratio in \eref{MCMCacceptratio} and sampling can be performed in $O(1)$.  There is only a small chance of moving to partitions with large enough elements to make the ratio more complex, as noted in \sref{largesampling} for elements with $k_i>7$. This should not suffice to increase the average complexity.  However for completeness consider the most extreme case when the partition $[n]$ is split into $[n/2,n/2]$ assuming $n$ is even.  The ratio in the acceptance probability is then
\begin{equation}
\frac{a_{[n/2,n/2]}}{a_{[n]}} = \binom{n}{\frac{n}{2}} 2^{\frac{n^2}{4}} ,
\end{equation}
involving large numbers but clearly larger than 1.  The reverse move which is based on the inverse can however be performed by iterative sampling as follows. Repeatedly draw from a Bernoulli$(\frac{1}{2})$ and reject the move as soon as 0 is drawn. In the unlikely event that a 1 is drawn a total of $n^2/4$ times consecutively the procedure moves through the binomial terms.  More Bernoulli random variables are then sampled with rate ranging from $1/n$ to $n/(n+2)$, but stopping and effectively rejecting as soon as a zero is drawn. A move may in principle take up to $n^2$ steps, but it is on average rejected in just 2 steps.   Even in this extreme case the move is still $O(1)$ on average leading to the conclusion that each move of the chain can be performed with no complexity overhead. 

\subsection{Speed of convergence}

The chain requires $(n-1)$ steps to become irreducible. Hence the speed of convergence is related to the eigenvalues of the transition matrix $U=T^{n-1}$ with the elements of $T$ corresponding to a single move of the MCMC chain.  The difference between the maximum and minimum element of each column of $U^k$ again decays exponentially with a rate $\alpha$ dependent on the largest eigenvalue of $U$ below 1.  Since the partition $[n]$ only has one DAG, $k$ of the order $n^2/\alpha$ is again needed for convergence on the relevant scale as in \sref{markovchainmethod}.  To explore how $\alpha$ depends on $n$ consider the explicit values for $n=2,\ldots,5$, which are found to be $\alpha=1.386,0.962,0.872,0.811$ respectively.  These decay with increasing $n$, but $\alpha\log(n)$ increases suggesting that convergence is attained for $k$ of the order of $n^2\log(n)$.

Since $(n-1)$ steps are required to first obtain $U$ and each step takes $O(1)$ the whole algorithm is $O(n^3\log(n))$.  Each time a partition is sampled a corresponding DAG can also be drawn according to \sref{reconstruction} without a complexity overhead.  By combining DAGs into partitions irreducibility is reached by shorter chains. Since there is also no need to check for acyclicity, as the samples are DAGs by construction, the resulting algorithm is much faster than the Markov chain method of \cite{mdb01} discussed in \sref{markovchainmethod}. Recall that the standard Markov chain only becomes irreducible in the time it takes for the hybrid partition method to converge.  The partition Markov chain can then be applied to sample reasonably large graphs while sharing some of the advantages of other Markov chain methods, such as the possibility of easily including restrictions.

Irreducibility could be achieved by an even shorter chain, for example by sampling from all partitions (or binary sequences) at each step.  The disadvantage lies in the potentially increased complexity of the acceptance probability evaluation and the associated sampling operation.  Such large moves would also reduce the probability of moving away from the most likely partition (with all $k_i=1$) increasing the chance that the chain gets temporarily stuck and so hindering the convergence.  Flipping a single element of the binary sequence allows more local and inexpensive moves with better acceptance probabilities.  Such moves should explore the space more easily which is why we opted for them here.

\section{Restrictions and variations}

A promising application of the Markov chain method, as particularly pursued in \cite{ic02,icr04}, is to impose restrictions on the DAGs to sample.  For example the number of edges and the degrees of the nodes can be restricted by simply rejecting any steps producing graphs which do not meet the requirements.  More complicated Markov chains are needed for sampling polytrees and DAGs with restricted density \citep{ic02,icr04} but the speed of convergence to uniformity remains an issue.  Although the space of restricted graphs is smaller, the probability of not moving for graphs on the border of the restricted set is increased due to the additional rejections.  If no recursive enumeration is possible for the restricted graphs, however, a Markov chain approach remains the only one available. For restrictions which can be incorporated combinatorially, direct enumeration again provides a more efficient method and we discuss some examples below. 

\subsection{Connected DAGs}

The main restriction imposed in \cite{ic02,mp04} is that DAGs be weakly connected (admitting  a path between every pair of nodes in the underlying undirected graph).  They were also counted by \cite{robinson73} and are recorded as sequence A082402 in \cite{sloanes}.  The standard Markov chain method can be simply modified so to only delete an arc if the resulting graph is also connected \citep{ic02} or, more efficiently, instead to reverse its direction otherwise \citep{mp04}. Checking for connectedness is of the order of the number of arcs making the algorithm $O(n^6)$. At each step either connectedness or acyclicity needs to be checked, the first when removing an edge and the second when adding one.  The resulting algorithm may than be actually slower then simply rejecting unconnected DAGs at the end. The fraction of connected DAGs tends to 1 as $n$ increases \citep{benderetal86,br88,robinson73};  for $n=4$ it is 82\% (for smaller $n$ we can enumerate the possibilities by hand), while it is $>99$\% and growing for $n>8$. Therefore the above modifications of the Markov chain, though theoretically interesting, are of no practical use, as it is much more efficient to sample a DAG at random and reject the disconnected ones.

When removing the $k$ outpoints of a connected DAG the rest breaks up into a set of smaller connected DAGs, hence a recursive combinatorial enumeration (involving a sum over the partitions of $m$) is possible.  Again the increase in algorithmic complexity makes a simple rejection step more efficient, apart from possibly for small $n$.

\subsection{Restricted new connections}

Because new outpoints are added at each stage of the recursive enumeration of DAGs, restrictions on the number of incoming arcs the previous nodes can receive from them are the simplest to treat.  Allowing up to a maximum $K$ new connections ($K\geq 1$ avoids only generating DAGs with no arcs) the recursions change to
\begin{equation} \label{atilderecur1}
\tilde{a}_{n,k}=\binom{n}{k}\sum_{s=1}^{m}\left[\sum_{i=1}^{\min\left(k,K\right)}\binom{k}{i}\right]^{s}\left[\sum_{i=0}^{\min\left(k,K\right)}\binom{k}{i}\right]^{m-s}\tilde{a}_{m,s} .
\end{equation}
For $K>k$ the above reduces to \eref{ankrecur}. A restriction on the minimum number of new connections can be incorporated analogously. Such restricted DAGs can easily be counted and then sampled as in \sref{enumeration}. In reconstructing the DAGs new arcs can be uniformly sampled from the binomial distributions in \eref{atilderecur1}.  Retaining the limit on the arcs received by the previous outpoints, it is also straightforward to separately limit the connections each new outpoint can send to the previous non-outpoints to a maximum of $\Kn$.  The corresponding DAGs can be sampled through
\begin{equation}
\tilde{a}_{n,k}=\binom{n}{k}\sum_{s=1}^{m}\left[\sum_{i=1}^{\min\left(k,K\right)}\binom{k}{i}\right]^{s}\left[\sum_{i=0}^{\min\left(m-s,\Kn\right)}\binom{m-s}{i}\right]^{k}\tilde{a}_{m,s} .
\end{equation}

\subsection{Restricted number of children}

Because they must be connected to, including previous outpoints when limiting the outgoing arcs from each new one is more complicated. Since arcs only originate from newly added outpoints, this means limiting the children to a certain $K$.  Denote by $C(k,m,s,K)$ the number of ways of adding $k$ new outpoints to a DAG with $m$ nodes and $s$ outpoints while restricting the number of children to $K$ or below. An expression for the ways $C(k,m,s,K)$ of linking $k$ new outpoints may be found by subtracting from the ways of connecting $k$ nodes to $m$ (with up to $K$ arcs each) those which leave any of the $s$ outpoints unconnected to 
\begin{eqnarray} \label{Crecur}
C(k,m,s,K)&=&\left[\sum_{i=0}^{\min\left(m,K\right)}\binom{m}{i}\right]^{k} \\
\nonumber & & {} - \sum_{i=1}^{s}\binom{s}{i}C(k,m-i,s-i,K), \\
\tilde{a}_{n,k} &=&\binom{n}{k}\sum_{s=1}^{m}C(k,m,s,K)\tilde{a}_{m,s} .
\end{eqnarray}
The graph is reconstructed by drawing from the $C(k,m,s,K)$ possibilities at each step,  and formula \eref{Crecur} simply suggests rejecting the cases which leave some of the $s$ outpoints with no incoming link. It is however more efficient to first derive the distribution of the number of arcs linking each new outpoint to the old ones when excluding the cases with a total $<s$. Samples from the resulting distribution where any old outpoints do not receive arcs are then rejected.  Arcs to the remaining nodes can be drawn subsequently.

For $C(k,m,s,K)$ much smaller than the first term on the right of \eref{Crecur}, or when the average number of new arcs linking to the old outpoints falls, the acceptance ratio also drops.  In this limit it may be convenient to first draw exactly one link to each of them and then sample the remaining arcs among all the nodes, including the $s$ old outpoints. The configurations where each receives $l_i$ incoming arcs are overcounted by a factor $F=\prod_{i=1}^{s}l_i$, so the sample should only be accepted with probability $1/F$.  

A balance of the two approaches may help minimise the rejection probability, but this example highlights that just knowing the number of possibilities is not sufficient for reconstruction, and a constructive method to generate them is preferable. A uniformly drawn DAG with a limit on the number of parents can be simply obtained by inverting all the arcs of a DAG with a limited number of children.

\subsection{Hybrid MCMC method to restrict parents}

As illustrated in the previous subsection, the combinatorial enumeration based on recursively removing outpoints reaches its limits when aiming to restrict the number of children or parents. The difficulties can be avoided by working partitionwise, rather than directly on individual DAGs.  Given a partition $[k_I, \ldots k_{i+1}, k_{i}, \ldots ,k_1]$ the number of ways, $G_{i,K}$, in which each of the $k_{i+1}$ nodes can receive up to $K$ edges from the $k_i$ in the next partition element and the remaining $S_{i-1}$ further up the DAG can be easily computed
\begin{equation}
G_{i,K}=\sum_{l=1}^{\min(K,k_i)}\binom{k_i}{l}\left[ \sum_{j=0}^{\min(S_{i-1}, K-l)}\binom{S_{i-1}}{j}\right] ,
\end{equation}
For $K=\infty$ this directly reduces to the number of edges in unrestricted DAGs: $G_{i,\infty}=(2^{k_i}-1) 2^{S_{i-1}}$.

The corresponding number of restricted DAGs for each partition is then simply
\begin{equation} \label{restrictedpartitiondagnumbers}
\tilde{a}_{[k_I,\ldots,k_1]} = \frac{n!}{k_I!\ldots k_1!} \prod_{i=1}^{I-1} (G_{i,K})^{k_{i+1}} .
\end{equation}
This can replace expression \eqref{partitiondagnumbers} in the acceptance ratio \eqref{accratio} of the MCMC method in \sref{hybridMCMC} providing a way to uniformly sample DAGs with a restricted number of parents.

\section{Weighted sample}

When generating DAGs by sampling triangular matrices their sparsity can be easily influenced by drawing the entries from a Bernoulli with $p \neq 1/2$. The associated weighting however acts on an underlying non-uniform space of DAGs. A weighted sample from an underlying uniform distribution may be preferable in practice when DAGs rather than triangular matrices are the object of study. The probability of a DAG with $l$ arcs should basically be proportional to $p^l(1-p)^{L-l}$ and analogous factors should be included in the recursive enumeration. Each of the $m-s$ non-outpoints may receive no links from the $k$ new ones with weight $(1-p)^k$, one link with weight $kp(1-p)^{k-1}$ and so on to include all the terms in the binomial expansion of $[(1-p)+p]^k=1$. The possibility that no arcs link to the $s$ old outpoints is excluded. The $k$ new outpoints on the other hand could be theoretically connected by $\binom{k}{2}$ links, but no arcs are allowed between them, leading to the recursion 
\begin{equation} \label{weightedahatnkrecur}
\hat{a}_{n,k}=\binom{n}{k}(1-p)^{\binom{k}{2}}\sum_{s=1}^{m}\left(1-(1-p)^k\right)^{s}\hat{a}_{m,s} ,
\end{equation}
where now $\hat{a}_{n,n}=(1-p)^{L}$ correspond to the DAGs with no arcs out of $L$ possibilities. It is chosen to set $0^0=1$ so that $\hat{a}_{1,1}=1$ even for $p=1$ where only one new outpoint is added at each step.  Weighted DAGs can be generated as in \sref{enumeration} where links are now added at each step with probability $p$.  The terms in the recursions are no longer integers in general (though for rational $p$ the numerator and denominator could be stored as a pair of integers) and exact sampling can be obtained up to the precision limits of the computational implementation.  

If many samples were required for a particular value of $p$ the limiting distribution could be derived again and used for sampling large graphs, as discussed in \sref{largesampling} for the case $p=1/2$.  However, as $p$ is reduced more weight is shifted towards DAGs with a larger number of outpoints added at each step and more terms would need to be included.  Often sparse DAGs are required in practice for modelling reasons where $p$ might scale like $1/n$ to keep the average number of parents low.  In the limit of small $p$ it may be necessary to calculate all the numbers in \eref{weightedahatnkrecur} up to a given precision.

The hybrid MCMC method of \sref{hybridMCMC} can be used instead to treat weighted DAGs.  Analogously to \eref{weightedahatnkrecur}, the weight of DAGs corresponding to a partition $P=[k_I,\ldots,k_1]$ can be expressed as
\begin{equation} 
\hat{a}_{[k_I,\ldots,k_1]}=\frac{n!}{k_I!\ldots k_1!} \prod_{i=1}^{I} (1-p)^{\binom{k_i}{2}} \prod_{i=1}^{I-1} \left(1-(1-p)^{k_{i}}\right)^{k_{i+1}} .
\end{equation}
The acceptance ratio of a proposed partition $\hat{P}$ where we split element $k_i$ from $P$ into two elements of size $c$ and $(k_{i}-c)$ simplifies to
\begin{eqnarray} \label{weightedMCMCacceptratio}
\frac{\hat{a}_{\hat{P}}}{\hat{a}_P} &=& \binom{k_i}{c}\left(\frac{1-(1-p)^{c}}{1-(1-p)^{k_i}}\right)^{k_{i+1}} \\ 
\nonumber & & {} \times \left(\frac{1-(1-p)2^{k_i-c}}{1-(1-p)^{k_{i-1}} \mathrm{I}_{\{ i>1 \}} }\right)^{c} (1-p)^{-c(k_i-c)} .
\end{eqnarray}
For $p=1/2$ this reduces to the expression in \eref{MCMCacceptratio}.  The hybrid MCMC method then allows us to efficiently sample DAGs weighted by their number of arcs without needing to calculate and store all the numbers in \eref{weightedahatnkrecur}.  For perfect sampling with rational $p$ the time and space complexity are at least $O(n^5)$ and $O(n^4)$ respectively as discussed in \sref{enumerationcomplexity} for uniform sampling.  The hybrid MCMC method therefore offers an alternative to sampling triangular matrices while removing the bias of operating on the wrong underlying space.

\section{Conclusions}

The recursive method for generating DAGs we analysed has the advantage over the highly adopted strategy of sampling triangular adjacency matrices of producing a uniformly distributed sample and over Markov chain based algorithms of avoiding convergence problems and being computationally more efficient.  The recursive method provides a perfectly uniform DAG on $n$ vertices with a setup time of $O(n^5)$ and a sampling time of $O(n^3)$ compared to the Markov chain method which provides an approximately uniform DAG in $O(n^5\log(n))$.  Comparing the actual computational time for $n=100$, the recursive method provides uniform DAGs thousands of times faster than the Markov chain.  Moreover, an approximately uniform DAG can instead be obtained through the limiting behaviour of the recursive enumeration in just $O(n^2)$.  The accuracy can be chosen at will and this minimal level of complexity means that the very large graphs which may be needed for practical applications can be generated. 

Along with using the limiting behaviour, we developed a novel hybrid MCMC setup based on the space of partitions of $n$.  Each partition is simply scored according to the number of DAGs belonging to it.  This offers the advantage compared to current alternatives of shortening the chain needed to reach irreducibility and avoiding the costs of checking for acyclicity.  Approximately uniform DAGs can be sampled in $O(n^3\log(n))$ which is a factor of $n^2$ smaller than current Markov chain methods.  Such an MCMC on the space of partitions can also be easily adapted to uniformly sample restricted DAGs, for example by including a restriction on the number of parents for each node or by weighting the edges.

The ideas discussed here could be integrated within MCMC schemes in the inference of Bayesian graphical models. A procedure can be built following the combinatorial construction where moves are suggested by going back a certain number of steps in the sequence of adding outpoints and resampling just the new connections at that step. As they are sampled uniformly, such moves are reversible. To ensure full space reachability, one could introduce some probability of proposing a move uniformly amongst all DAGs or according to the standard MCMC on network structures \citep{art:MadiganY95}.  Alternatively, a number of moves through partitions can be defined based on the underlying combinatorial structure.

The hybrid MCMC method based on partitions is similar in spirit to the sampling in the space of orders of \cite{art:FriedmanK2003} which actually acts on the space of triangular matrices.  Scoring all the DAGs in a given partition therefore possibly has similar potential to greatly improve the mixing properties of the chain produced, with the advantage of avoiding the bias inherent in the order version of acting on a different space.  The local changes in partition elements could also possibly propose structures which are in some sense close in the space of graphs while other moves based on the combinatorial structure might be tuned to the local scores.  Since neither the original MCMC on network structures nor the order version appear to be completely satisfactory in practice \citep{art:GrzegorczykH2008} the possibilities outlined above seem promising.

Interpreted as Bayesian networks with random variables placed on each node, DAGs encode conditional independencies of multivariate distributions.  However, several DAGs might encode he same set of conditional independencies, meaning that they typically cannot be uniquely identified from data.  The set of DAGs with the same probability distribution, also known as the Markov equivalence class, can be represented by an essential graph \citep{amp97} or a completed partially DAG.  In practice, it might be desirable to sample equivalence classes and work on the space of essential graphs as opposed to DAGs.  However, although essential graphs can now be counted, the current formula takes exponential time and their number is therefore only known up to $n=13$ \citep{steinsky13}.  Previously, essential graphs had only been counted by running through all possible DAGs for $n\leq 10$ \citep{gp02} with the $n=10$ term corrected in sequence A007984 of \cite{sloanes}.  

MCMC methods can be restricted to the space of essential graphs, for example an implementation of the general framework of \cite{art:MadiganY95} was given in \cite{madiganetal96} while a Markov chain algorithm specifically targeted at uniform sampling was developed by \cite{pena07}.  Comparing to the Markov chain for sampling DAGs uniformly \citep{mdb01,scutari10} the algorithm for essential graphs is notably slower, taking a similar time to make each step for an essential graph with about 20 or 30 nodes as for a DAG with 100.  Although no convergence details are given, only a very small fraction of the proposed moves result in an essential graph suggesting that the convergence would also be much slower than for DAGs.

Although little is known about the exact number of essential graphs beyond the results in \cite{steinsky13}, the number of essential graphs which are also DAGs is known \citep{steinsky03}.  More importantly, the asymptotic ratio of the number of DAGs to those which are essential graphs is under 13.7 \citep{steinsky04}.  Since there are more essential graphs than just those which are also DAGs, 13.7 is a bound on the average number of DAGs corresponding to each essential graph, while the actual ratio seems to be less than 4.  With such a low bound, it might be worth thinking of rejection or importance sampling strategies, starting from the uniform DAG generation discussed here, possibly leading to a simple and more efficient alternative for sampling essential graphs.

\appendix

\section{Convergence of the Markov chain sampler} \label{markovconvergence}

To explore the convergence of the Markov chain sampler, consider the spectral decomposition of the real, symmetric transition matrix $T$
\begin{equation} \label{spectraldecomposition}
 T = \sum_{i=1}^{a_n} \lambda_i v_i v_i' ,
\end{equation}
in terms of its real eigenvalues $\lambda$, which can be labelled by their size
\begin{equation}
 1=\lambda_1 > \vert \lambda_2 \vert \geq \vert \lambda_3 \vert \geq \ldots \geq \vert \lambda_{a_n} \vert,
\end{equation}
and orthonormal eigenvectors $v$.  The vector $v_1$ is simply $(1,\ldots,1)'/\sqrt{a_n}$ and responsible for the underlying uniform distribution.  The matrix 
\begin{equation} \label{Tjdecomp}
T^j = v_1v_1' + \sum_{i=2}^{a_n} \lambda_i^j v_i v_i' ,
\end{equation}
then converges to this uniform background with a rate depending on the remaining eigenvalues and $\lambda_2$ in particular.  This is easiest to see in terms of the Frobenius norm, which, setting $S_j = T^j - v_1v_1' $ to be the transition matrix with the uniform background removed, satisfies
\begin{equation}
\vert \vert S_j \vert \vert = \sqrt{\sum_{m,l=1}^{a_n}(S_j)_{ml}^2} = \sqrt{\sum_{i=2}^{a_n}\lambda_i^{2j}} \leq \vert\lambda_2\vert^{j} \sqrt{a_n-1} = \vert\lambda_2\vert^{j} \vert\vert S_0 \vert \vert ,
\end{equation}
as $T^0$ is the identity matrix.  From the resulting inequality,
\begin{equation}
\vert\vert T^j - v_1v_1' \vert\vert \leq \vert \lambda_2 \vert ^j \vert\vert I - v_1v_1' \vert\vert ,
\end{equation}
since the Frobenius norm involves a sum over the elements squared, it follows that every element of $T^j$ must approach the uniform background exponentially at least as fast as $\sim \exp(j \log\vert\lambda_2\vert)$, or
\begin{equation} \label{Tjmllim}
\vert (T^j)_{ml} - \frac{1}{a_n}\vert \leq  C_{ml} \exp(j \log\vert\lambda_2\vert)
\end{equation}
for some constants $C_{ml}$.  A similar inequality likewise holds for the maximum and minimum elements of $T^j$ and their difference with corresponding constants.  We can obtain upper bounds for the constants by returning to \eref{Tjdecomp}.
\begin{equation}
\vert (T^j)_{ml} - \frac{1}{a_n}\vert =  \vert \sum_{i=2}^{a_n} \lambda_i^j (v_i)_m (v_i)_l  \vert \leq 
\vert \lambda_2\vert^j \sum_{i=2}^{a_n} \vert (v_i)_m (v_i)_l  \vert \leq \vert \lambda_2\vert^j (a_n-1)
\end{equation}
The comparison to \eref{Tjmllim} directly gives $C_{ml} < a_n$.

For the irreducible matrix $U=T^{2L}$, its powers $U^k$ converge to uniformity $\sim\exp(-\alpha k)$ with a rate given by
\begin{equation} \label{alphadef}
\alpha \geq -2L\log\vert\lambda_2\vert
\end{equation}
which is simply minus the log of the largest eigenvalue of $U$ below the one at unity.  For the difference from the uniform background to converge on the scale below $1/a_n$, as discussed in \sref{nonuniformitychain}, \eref{Tjmllim} provides an upper bound that $-j\log\vert\lambda_2\vert$, or equivalently $-2Lk\log\vert\lambda_2\vert$, be of order $n^2$. 

For comparison with other methods, and in particular the enumeration method studied in this paper, it would be useful to obtain a tight lower bound for convergence on the scale well below $1/a_n$.  If $v_2$, like $v_1$, also had its weight evenly spread among all $a_n$ DAGs, for example if $v_2$ were $(\pm 1,\ldots, \pm 1)'/\sqrt{a_n}$, then the term $\lambda_2 v_2 v_2'$ would start at the scale of $1/a_n$ and converge directly for $-j\log\vert\lambda_2\vert$ of order $1$.  The overall convergence would depend on the smaller eigenvalues and how the weight of their eigenvectors is spread amongst the DAGs.  For a better handle on this, we can focus on the diagonal elements of the transition matrix in \eref{spectraldecomposition}, which we can write as
\begin{equation}
 \mathrm{diag} (T) = \sum_{i=1}^{a_n} \lambda_i D^{i}, \qquad D^{i} = \mathrm{diag} \left(v_i v_i'\right) ,
\end{equation}
where the elements of the diagonal matrices $D^i$ are real, positive and satisfy
\begin{equation} \label{diagrestrictions}
\sum_{i=1}^{a_n} D^{j}_{i,i} = 1, \qquad \sum_{i=1}^{a_n} D^{i}_{j,j} = 1 .
\end{equation}

The diagonal elements of $T$ depend on the number of edges in each DAG and how they are arranged, but we can consider the two extremes.  For the empty DAG with no edges, the probability of staying still is $1/n$, so if this DAG is chosen as the first element of the transition matrix
\begin{equation}
\sum_{i=1}^{a_n}\lambda_i D^{i}_{1,1} =\frac{1}{n} .
\end{equation}
At the other end of the spectrum are all the $n!$ DAGs with $L$ arcs and a staying probability of $1/2+1/2n$.  If one of them is chosen as the last element of the transition matrix
\begin{equation}
\sum_{i=1}^{a_n}\lambda_i D^{i}_{a_n,a_n} = \frac{1}{2} +\frac{1}{2n} .
\end{equation}
Assuming the eigenvalues are all positive, this along with \eref{diagrestrictions} implies that the eigenvalues must cover a range from $O(1)$ to $O(1/n)$ and that the DAG with no edges must have its weight (in the matrices $D^{i}$) preferentially spread over the smaller eigenvalues.  Similarly, the DAGs with as many edges as possible concentrate their weight over the larger eigenvalues.  Intuitively, the different staying probabilities of the DAGs is encoded in how far down the chain of eigenvalues the diagonal contribution is stored.

In fact, looking at the examples for $n=2$ and $n=3$ for which we can fill out the transition matrix, we find that matrices are positive definite and that we have repeated eigenvalues for $n=3$.  Looking at the sum over the largest eigenvalues below 1,
\begin{equation}
\sum_{i}D^i \delta_{\lambda_2,\lambda_i} ,
\end{equation}
we find that almost all of the weight is evenly spread amongst the $n!$ DAGs with the highest number of edges.  This suggests that the corresponding diagonal elements of the transition matrix are bounded below by
\begin{equation}
(T^{j})_{a_n,a_n} -\frac{1}{a_n} \gtrsim \frac{\vert\lambda_2\vert^j}{n!} .
\end{equation}
Convergence on the required scale still requires $-j\log\vert\lambda_2\vert$ to be order $n^2$ to get the right hand side below $1/a_n$.  When moving to the irreducible matrix $U=T^{2L}$, we have $\alpha = -2L \log\vert\lambda_2\vert$ when combined with \eref{alphadef}, making $j$ of the order of $n^4/\alpha$, or $k$ of the order $n^2/\alpha$ as in \sref{markovchainmethod}.

\section{Complexity of the enumeration method} \label{complexity}

The time complexity of computing all the integers $a_{j,k}$, $b_{j,k}$ as well as the totals $a_j$ for $j$ up to $n$ is now considered. A binary representation is used to treat the integers, and from \eref{totalDAG} it follows that the number of bits needed to store $a_j$ grows like $\binom{j}{2}$ or is of order $j^2$.  Once all the $a_{j,k}$ for all $j$ up to $n-1$ have been calculated, the $b_{n,k}$ can be computed using \eref{ankrecur} in the following way.  For $k>1$, for each $s$ first multiply $a_{m,s}$ by $(2^k-1)^s$.  This can be done in $s$ steps where first a simple shift of the binary representation is performed to multiply by $2^k$. Then a copy of the previous number is subtracted, an operation which takes a time proportional to the length of its binary representation which is bounded by $n^2$.  For each $s$, calculating the term in the sum takes $O(sn^2)$ while the final multiplication by $2^{k(m-s)}$ is again just a shift of the binary sequence.  Finding all the terms in the sum is then $O(n^3)$.  Adding the terms to obtain $b_{n,k}$ then means adding up to $n$ sequences of length up to $n^2$ which is also $O(n^3)$.

Next the $a_{n,k}$ are obtained by multiplying by the binomial coefficients.  These can be calculated recursively without any complexity overhead over recursively calculating $b_{j,k}$.  The binomial coefficients also have a binary length bounded by $n$ so that multiplying $b_{n,k}$ by $\binom{n}{k}$ is still below $O(n^3)$.  However the $b_{n,k}$ need to be calculated for all $1\leq k\leq n$, which leads to a complexity of $O(n^4)$ for computing $a_{n,k}$ and $b_{n,k}$ given all the previous values.  Finding $a_n$ by summing the $a_{n,k}$ is also order $n^3$ so it does not further affect the complexity.

Computing all the integers $a_{j,k}$, $b_{j,k}$ as well as the totals $a_j$ for $j$ up to $n$ then involves repeating the above process $n$ times giving a final complexity of $O(n^5)$.

The above steps provide an upper bound for completing the first step of the algorithm in \sref{enumeration} but it uses the assumption that all of the $a_{n,k}$ have a similar length to $a_n$.  As seen in \sref{largesampling} though, $a_{n,k}$ has a similar length to $a_n$ only for a limited number of $k$ and the length then decays to 1 for $a_{n,n}$.  Looking at the distribution of $s\log(a_{m,s})$ can then give a more accurate estimate of the complexity of finding all the terms in the sum to calculate $b_{n,k}$.  This also gives a lower bound since at each of the $s$ steps in the simplified algorithm above the length of the binary representation is actually increased.  Numerically, this distribution has a single peak for $s$ just beyond $m/2$ but its maximal value seems to grow like $n^3$ which also leads to a total complexity of $O(n^5)$.

When sampling the DAGs, the first step of finding $k$ given an integer between 1 and $a_n$ involves subtracting up to $n$ numbers of binary length up to $n^2$ and is $O(n^3)$.  Then given $(n,k)$ we look to sample $(m,s)$ again using the sums over $s$ that appear in \eref{ankrecur}.  As discussed above, performing the sum is $O(n^3)$ while in the end the sampling is performed up to $n$ times which seems to give a total complexity of $O(n^4)$.  However, the sum of the outpoints sampled has to be exactly $n$ while the complexity of sampling each $k_i$ is bounded by $(k_i+1-\delta_{k_i,1})n^2$.  With $\sum_{i}k_i = n$, then $\sum_{i}(k_i+1)n^2 \leq 2 n^3$ which reduces the total complexity to $O(n^3)$.  Also, since there is effectively no chance of choosing a large $k_i$ as discussed in \sref{largesampling} the complexity of sampling $k_1$ and each of the following $k_i$ reduces to $O(n^2)$ immediately leading to a total complexity of $O(n^3)$.

\section{Pseudocode for uniform DAG sampling} \label{pseudocode}

The code uses arbitrary precision integers as in Maple or as provided by the GMP library and the `bigz' package in R.  First we recursively calculate and store the numbers $a_{n,k}$ and $b_{n,k}$ for $n\leq N$
\begin{algorithmic}
\For{$n=1$ to $N$}
  \For{$k=1$ to $n-1$} \Comment{when $n>1$}
    \State $b_{n,k} \gets \sum_{s=1}^{n-k}(2^k-1)^{s}2^{k(n-k-s)}a_{n-k,s}$
    \State $a_{n,k} \gets \binom{n}{k}b_{n,k}$
  \EndFor
  \State $a_{n,n}\gets 1, \qquad b_{n,n}\gets 1$
  \State $a_{n} \gets \sum_{k=1}^{n}a_{n,k}$
\EndFor
\end{algorithmic}
If the binomial coefficients are not readily available they can be built recursively in the same loops with no computational overhead. Next we sample an integer between 1 and $a_n$
\begin{algorithmic}
\State $\mathrm{numbits} \gets \lceil \log_{2}(a_n)\rceil$ 
\Repeat
  \State $r\gets 1$
  \For{$j=1$ to $\mathrm{numbits}$}
    \State $r\gets r + 2^{j-1}$rand$\{0,1\}$
  \EndFor
\Until $r\leq a_n$
\end{algorithmic}
where `rand$\{0,1\}$' provides a 0 or a 1 with probability $1/2$.  Now we use the integer to sample the number of outpoints $k$
\begin{algorithmic}
\State $k \gets 1$ 
\While{$r>a_{n,k}$}
  \State $r\gets r - a_{n,k}$ 
  \State $k\gets k+1$
\EndWhile
\State $i\gets 1$
\State $\vec{k}_{i} \gets k$
\end{algorithmic}
which we store as the first element of a vector $\vec{k}$. The current value of $r$ should be between 1 and $a_{n,k}$ which we rescale to between 1 and $b_{n,k}$
\begin{algorithmic}
\State $r \gets \lceil \frac{r}{\binom{n}{k}}\rceil$ 
\State $m\gets n-k$
\end{algorithmic}
Next we recursively generate the outpoints in the loop
\begin{algorithmic}
\While{$m>0$}
  \State $s\gets1$
  \State $t\gets (2^k-1)^{s}2^{k(m-s)}a_{m,s}$
  \While{$r>t$}
    \State $r\gets r - t, \qquad s\gets s+1$
    \State $t\gets (2^k-1)^{s}2^{k(m-s)}a_{m,s}$
  \EndWhile
  \State $r \gets \lceil \frac{r}{\binom{m}{s}(2^k-1)^{s}2^{k(m-s)}}\rceil$ 
  \State $n\gets m, \qquad k \gets s$
  \State $m \gets n-k$
  \State $i\gets i+1$
  \State $\vec{k}_{i} \gets k$
\EndWhile
\State $I\gets i$
\end{algorithmic}
The resulting vector $\vec{k}$ should be of length $I$ and have its elements sum to $n$.  We can now use this to fill the lower triangle of an empty matrix $Q$
\begin{algorithmic}
\State $j\gets \vec{k}_{I}$
\For{$i=I$ to $2$} \Comment{when $I>1$}
  \For{$l=j-\vec{k}_{i}+1$ to $j$}
    \Repeat
      \For{$m=j+1$ to $j+\vec{k}_{i-1}$}
        \State $Q_{m,l} \gets$ rand$\{0,1\}$
      \EndFor 
    \Until $\sum_{m=j+1}^{j+\vec{k}_{i-1}}Q_{m,l} > 0$
    \For{$m=j+\vec{k}_{i-1}+1$ to $n$} \Comment{when $i>2$}
      \State $Q_{m,l} \gets$ rand$\{0,1\}$
    \EndFor
  \EndFor
  \State $j \gets j+\vec{k}_{i-1}$
\EndFor
\end{algorithmic}
Finally we sample a permutation $\pi$ by randomly drawing all the integers $\{1,\ldots,n\}$ without replacement.  To obtain the adjacency matrix $R$ of the uniformly sampled DAG we correspondingly permute the column and row labels of $Q$ via $R_{\pi(m),\pi(l)}=Q_{m,l}$.

\section{Limiting conditional outpoint distribution} \label{limitcondprobsderivation}

In the expression defining $b_{n,k}$ in \eref{ankrecur}, we can reorganise the powers of two
\begin{equation}
b_{n,k}=2^{km}a_m\sum_{s=1}^{m}\left(1-\frac{1}{2^k}\right)^{s} \frac{a_{m,s}}{a_m} , 
\end{equation}
and artificially bring out a factor of $a_m$.  For large $m$, the fraction $a_{m,s}/a_m$ can be replaced by its limit $A_s$ since it is only non-zero for a small number of $s$ at a given accuracy 
\begin{equation} \label{bnkpropeqn}
b_{n,k}\propto \sum_{s}\left(1-\frac{1}{2^k}\right)^{s} A_s . 
\end{equation}
Given $k$, to sample the next number of outpoints $s$ we can sample uniformly between 1 and $b_{n,k}$ as in \sref{outpoints}.  The limiting probability of sampling each value of $s$ is then 
\begin{equation}
P(s\mid k) \propto \left(1-\frac{1}{2^k}\right)^{s} A_s, 
\end{equation}
which through normalisation reduces to \eref{limitcondprobs}.


\bibliographystyle{spbasic}      
\bibliography{urgladbib}   

\end{document}